\documentstyle[12pt]{article}
\begin{document}
\title{The Nonradial Oscillation Node Precession of Neutron Stars} 
\author{{Haochen Li}\\
   {\small Physics Department, Washington University}\\
   {\small  St. Louis, MO 63143}}
\date{March 28, 2001}
\maketitle
 
\begin{abstract}                       
The standing wave nodes of nonradial oscillations on a neutron star crust will drift with a definite angle velocity around rotational pole due to the rotation of neutron stars. This is called the nonradial oscillation node precession of neutron stars. This article estimated the precession velocity and pointed out that it merely lies on the star's rotation velocity and the angular order of spherical harmonic $l$ by one order approximation. If we suppose that oscillations effect the particles' escaping from the polar cap of a neutron star, so that the antinode and node areas of the standing waves have different radiative intensity, several unusual conclusions are acquired by reviewing the observation of pulsars which had already been taken as neutron stars. For example, the drifting subpulse period $P_{3}$ can be gotten from the width of subpulses and order $l$; the larger velocity drift may produce the peak structure of average pulse profiles; the dissimilar radiation phenomena between neighboring periods generated from drift provide a reasonable explanation of interpulses which have been found on some pulsars. 
\end{abstract}

\clearpage 

\begin{center} 
\section{Introduction}
\end{center}

\paragraph{}Boriakoff (1976) had detected quasi-periodic micropulsations within the subpulses of PSR 2016+28, and inclined to take it as nonradial oscillations of neutron stars. In the pulsar polar cap model (Radhakrishnan and Cooke 1969) the radio pulse is produced by the coherent radiation of particles escaping from a certain surface area of the star(polar cap)along the magnetic field lines. Because of the high particle velocity, the radiation is emitted in a narrow cone, the axis of which coincide with the velocity vector of particles, which is tangential to the magnetic field lines. Since these are periodically distorted by the star's vibration, the radiation cone will periodically change directions, switching on and off the radiopulse illumination of the observer(modulation). Van Horn (1980) pointed out that rotating, magnetized neutron stars can support a rich variety of oscillation modes and firstly suggested a possible association of subpulse drift and torsional oscillations. The special terms of the neutron stars are considered for calculating one order approximation of the frequency split of the torsional oscillations in Section 2. And using this result, we will discuss phenomena such as drifting subpulses, average pulse profiles and interpulses of pulsars in Section 3. Section 4 is the summary.

\begin{center}
\section{Theory of Neutron Star Oscillation Node Precession}
\end{center} 

\paragraph{}Ruderman (1968) firstly pointed out torsional oscillation modes of neutron star crusts. Hansen and Cioffi (1980) calculated the periods of those for a range of stellar models and found those associated with fundamental modes have periods of around 20 ms. We can use this result to estimate the lowest frequency of torsional oscillation of neutron stars as$$\omega_{0}={\frac{2\pi}{20ms}}=100\pi s^{-1}.$$ 
The rotation angular velocity of neutron star  can be considered as $2\pi s^{-1}$,  thus the ratio is $$\epsilon={\frac{\Omega}{\omega_{0}}}=0.02.$$
We can see that although the angular velocity of neutron star rotation is much larger than that of common stars, it is still small compared with the frequency of self-oscillation. This inspires us that the oscillation node precession theory which has been established on other heavenly bodies can be used on neutron stars (also because the torsional oscillation is little sensitive to sphere models, see Van Horn 1980, and its results is simple). That is, we can take rotation effect as perturbation to solve the sphere oscillation equations just as Ledoux (1951) did on gaseous stars and MacDonald and Ness (1961) did on the earth, and the frequency of free oscillation of sphere crust is the sum of the undisturbed frequency plus the perturbation frequency:$$\omega=\omega_{0}+\omega^{1}.$$
As one order approximation for torsional oscillation,$$\omega^{1}={\frac{m}{l(l+1)}}\Omega,$$
where $l$ and $m$ are integers denoting angular orders of spherical harmonic.
As the theory of oscillation of stars (Ledoux 1951) and the earth(MacDonald and Ness 1961) has noted, each value of $m$ has two travelling waves associated with it. In the case of the earth one wave travels eastward, and its rate of travel is decreased by the angular velocity of earth rotation; the other travels westward, and its rate is faster. The waves corresponded with neighboring values of $m$ have relative angular velocity $${\frac{\Omega}{l(l+1)}}.$$
The combined effect is to produce a standing-wave pattern that for a given value of $m$ moves westward with the angular velocity $${\frac{\Omega}{l(l+1)}}$$ of its nodes, which is well known in seismology as the node precession of oscillations. And this is just the result we will use next to recur to the attempt which Van Horn (1980) had made to connect the torsional oscillation of rotating neutron stars with the observation phenomena of pulsars.     

\begin{center}
\section{Discussion}

\subsection{Drifting Subpulses}
\end{center}
\paragraph{}
     We suppose that the node and antinode of the standing wave separately correspond with those of subpulse radiate wave pattern, i.e., drifting subpulses reflect the node precession. Then the degrees of subpulse drift in one period of a pulsar rotation (Manchester and Taylor 1977) is $$D_{\phi}={\frac{\Omega}{l(l+1)}} P_{1}={\frac{360}{l(l+1)}},$$
where $P_{1}$ is the pulsar rotational period and 360 of longitude are equal to one pulsar period. We can see that when $D_{\phi}$ is smaller than the width of subpulses (the drifting subpulse observation results are exactly so, see Manchester and Taylor 1977), then we get the subpulse drifting-band spacing $$P_{3}={\frac{\frac{P_{2}}{P_{1}}\times360}{D_{\phi}}}={\frac{l(l+1)}{\frac{P_{1}}{P_{2}}}}$$(in units of $P_{1}$),where $P_{2}$ is the subpulse period (converted from degrees of longitude). We calculated the values of $P_{3}$ for several pulsars using the observational data from Van Horn(1980) and Wright and Fowler(1981). The results are listed in Table 1. Note that these are acquired with larger values of $l$, with which the values of $P_{3}$ increase. Several proximal values have been enumerated in the table for compare. The difference between theoretical and observational values probably due to error and disconsidering of the coupling of several values of $l$. The different values of $P_{3}$ in one pulsar are deemed to mode switch of different $l$.


\begin{center}
\subsection{Average Pulses}
\end{center}

\paragraph{}Theoretically we have no reasons to believe that the drifting pace of subpulses is always small. Then, for convenience we define drifting rate as $$V={\frac{\frac{P_{1}}{P_{2}}}{l(l+1)}},$$        
which represents drift space (in units of $P_{2}$) in each rotational period of a pulsar. Because we could not find the integer drifting space (integer $V$), so the practically observed rate $V'$ rest with the decimal part of $V$. For example, if $V=3/2$ or $1/2$, then $V'=1/2$; if $V=5/3$, then $V'=2/3$ or $-1/3$(minus sign represents opposite drift direction). Here it implies that ${\frac{P_{1}}{P_{2}}}$ is integer which goes on the fact that it here represents the node number of the standing wave along the longitude of the sphere. When $l=1$ or $2$(the fundamental mode which the oscillation is most likely on), it is easy to determine that $V'$ will frequently get $1/2$, $1/6$, $2/6$(the same as $4/6$), etc. Unlike the smaller drifting pace discussed in Section 3.1, these values of $V'$ are too great to be detected as the drift we commonly mean (we do not know whether PSR 2303+30 listed in Table 1. belongs to these small $l$ modes). But in this situation subpulses will appear more frequently at the several fixures in the general radiate windows. The average pulse profiles imitated by computer program through adding a great many drifting periods show peak structures as displayed in Fig.1.


\begin{center} 
\subsection{Interpulses}
\end{center}
\paragraph{}If the pulses of pulsars can embody the node precession of standing waves around the longitude of neutron stars, then according to the observed drifting pace $V'$ discussed above, it must have the circumstances that neighboring periods of pulses have different observational pictures especially when the standing wave length is longer than the general radiate window. For instance, once we see a pulse which practically is a fraction of the standing wave length(a node or nearby), then next period we see the antinode or nearby($V'=1/2$ is very common), therefore the different intensity pulses alternately occur along with the integral periods of rotation. This can give an natural explanation of interpulses (Manchester and Taylor 1977). The weaker pulse will be surely inclined to the nearby node(or antinode) area which should have stronger radiation. That is why the degrees between neighboring pulses are not exactly 180(Manchester and Taylor 1977). If this is true, it means that the real periods of the interpulse pulsars are only half of those we believe now.
\begin{center}
\section{Summary}
\end{center}

\paragraph{}The sphere free oscillation has been proved by theory and  observation to be a very common phenomena in the world of stars and planets. Many prior works have supposed this happens on neutron stars which have so great density and so rapid rotation (Cheng and Ruderman 1980; Harding and Tademaru 1981; McDermott, Van Horn, and Hansen 1988; Cordes, Weisberg, and Hankins 1990). Although the mechanism of radiation affected by oscillation has not been clearly discussed ( which obviously is a very important problem), it gives a very natural explanation to drifting subpulse phenomena, the generalization of which reasonably gives clear pictures of pulsar's fundamental observation facts such as average pulse profile, interpulse, etc. The theory also has great potential in explanation of mode changing, micropulses, and glitches which maybe the author will discuss later. 
\paragraph{}It should be pointed out that although the theoretic values of $P_{3}$ we get in Section 3.1 using bigger $l$ are in good agreement with the observation values(see Table 1.), it does not mean that the actual oscillation orders are always high. The lower modes (small $l$) are not considered in the calculating of $P_{3}$ is because the larger scale drift devote little observational effect(this can be seen from the discussion of Section 3.2). Actually it is most probable that more than one mode of oscillation are simultaneously sustained on the star crust, and the observation phenomena is only the coupling of these modes. In mode switching, the dominant precession rate changes sequentially. The advanced job need to determine the relationship of $l$, $m$ and the width of subpulses, so that we can know exactly the parameters of the stars' oscillation and more detailed knowledge about a given pulsar. 
 
\paragraph{}I wish to thank Xinji Wu and Xiaofei Chen of Peking University for the helpful discussions.
\clearpage 

\clearpage 

Fig.1.   The sketch maps of average pulse formed by great pace drifting. The abscissa is longitude and the ordinate is intensity, and the numbers only have relative meaning. 1) is single pulse which indicates our suppose that there are only two gaussian subpulses in one general pulse window and this measures up the observational fact; 2)-3) are both average pulses with $V'=1$, distinct at the original positions; 4)-6) are $V'=1/2$, $2/6$, $1/20$ separately. The adding times are all $10^{4}$. It is obvious that the figures will keep stable on more adding times.

\clearpage 

        Table 1. The periods of drifting subpulses.
\begin{center}
\begin{tabular}{ccccccc} \hline \hline
PSR      & $P_{1}$(s) & $P_{2}$(ms) & ${\frac{P_{1}}{P_{2}}}$ & $l$ & $P_{3}$(Theory) & $P_{3}$(Observation) \\ \hline
         &            &             &                         &  19 &   18           &               \\
1944+17  &  0.440     &   21        &        21               &  20 &   20           &    20         \\
         &            &             &                         &  21 &   22           &               \\  \hline
         &            &             &                         &  8  &   4.2          &               \\
         &            &             &                         &  9  &   5.3          &               \\
         &            &             &                         &  10 &   6.5          &    4.5        \\
0031-07  &  0.943     &  55         &      17                 &  11 &   7.8          &    6.8        \\
         &            &             &                         &  12 &   9.2          &    12.5       \\
         &            &             &                         &  13 &   11           &               \\
         &            &             &                         &  14 &   12           &               \\
         &            &             &                         &  15 &   14           &               \\  \hline
         &            &             &                         &  8  &   1.7          &    2.11       \\
0943+10  &  1.097     &    26       &       42                &  9  &   2.1          &    or         \\
         &            &             &                         &  10 &   2.6          &    1.90       \\   \hline
         &            &             &                         &  16 &   10           &               \\
0809+74  &  1.292     &   50        &       26                &  17 &   12           &    11.0       \\
         &            &             &                         &  18 &   13           &               \\  \hline
         &            &             &                         &  18 &   3.8          &               \\
1919+21  &  1.337     &  15         &       89                &  19 &   4.3          &    4.2        \\
         &            &             &                         &  20 &   4.7          &               \\  \hline
         &            &             &                         &  18 &   5.9          &               \\
0301+19  &  1.387     &  24         &       58                &  19 &   6.6          &     6.4       \\
         &            &             &                         &  20 &   7.2          &               \\  \hline
         &            &             &                         &  13 &   1.7          &               \\
2303+30  &  1.575     &  15         &       105               &  14 &   2.0          & $\approx2$    \\
         &            &             &                         &  15 &   2.3          &               \\ \hline
         &            &             &                         &  8  &   2.14         &               \\
1237+25  &  1.38      &  41.0       &      33.7               &  9  &   2.67         & $2.8\pm0.1$   \\
         &            &             &                         &  10 &   3.26         &               \\ \hline

\end{tabular}

\end{center}

\end{document}